\documentclass[aip,jcp,amsmath,amsfonts,floatfix,reprint]{revtex4-1}

\usepackage{graphicx}
\usepackage{dcolumn}
\usepackage{bm}
\usepackage[utf8]{inputenc}
\usepackage{bbold}
\usepackage{color}
\usepackage[normalem]{ulem}

\begin{document}

\title{Charge susceptibility and conductances\\ of a double quantum dot}
\author{V.~Talbo}
\author{M.~Lavagna}
\email{mireille.lavagna@cea.fr}
\affiliation{Univ. Grenoble Alpes, CEA, INAC, PHELIQS, F-38000 Grenoble, France}
\author{T.Q.~Duong}
\author{A.~Cr\'epieux}
\affiliation{Aix Marseille Univ, Universit\'e de Toulon, CNRS, CPT UMR 7332, 13288 Marseille, France}

\begin{abstract}
We calculate the charge susceptibility and the linear and differential conductances of a double quantum dot coupled to two metallic reservoirs both at equilibrium and when the system is driven away from equilibrium. This work is motivated by recent progress in 
the realization of solid state spin qubits. The calculations are performed by using the Keldysh nonequilibrium Green function technique. In the noninteracting case, we give the analytical expression for the electrical current and deduce from there the linear 
conductance as a function of the gate voltages applied to the dots, leading to a characteristic charge stability diagram. We determine the charge susceptibility which also exhibits peaks as a function of gate voltages.  We show how the study can be extended to 
the case of an interacting quantum dot.
\end{abstract}

\maketitle

\section{Introduction}

The idea introduced two decades ago of a quantum computer based on solid state spin qubits\cite{Loss1998} has led to an intensive effort in the realization of spin qubits on the basis of double quantum dots\cite{Vaart1995,Fujisawa1998,Oosterkamp1998}. The 
basic idea is to manipulate the spin encoded in the first of the quantum dots by means of various dc or ac external fields, then use the quantum exchange interdot coupling to carry out two-qubit operations, and finally readout the information on the spin encoded 
in the second quantum dot. The challenge becomes all the more accessible now that long spin coherence time has been recently achieved for individual spin qubits\cite{Muhonen2014,Maurand2016} which ensures high-fidelity to quantum computation 
operations\cite{Veldhorst2014}. This quest for realizing solid state quantum bits has motivated parallel theoretical studies on double quantum dots. The electron-electron interactions when present have been taken into account in a capacitive model 
with an additional interdot capacitance. It has thus been possible to establish the charge stability diagram of these systems in which the Coulomb oscillations of conductance observed in a single quantum dot are changed into a characteristic 
honeycomb structure as a function of the gate voltages applied to each dot \cite{Matveev1996,Wiel2002}. Another topic that has been widely discussed in the last years on both experimental and theoretical sides, is the possibility of exposing the double quantum 
dot to an electromagnetic radiation (i.e. to an ac external field) allowing the transfer of an electron from one to the other reservoir even at zero bias voltage\cite{Hazelzet2001, Lopez2002, Golovach2004, Sanchez2006, Riwar2010, Cottet2011}. In this way, ac-
driven double quantum dots act as either charge or spin pumps. The transport can then be either incoherent via sequential tunneling processes or coherent via inelastic cotunnelling processes. Most of the theoretical studies so far have been done by using the 
master equations\cite{Golovach2004, Cottet2011} or real time diagrammatic approach\cite{Riwar2010} or time evolution of the density matrix\cite{Hazelzet2001,Sanchez2006}. It is worth noting that even if these methods make it possible to describe the regimes 
of either weak or strong interdot tunnel coupling, their domain of validity is mainly restricted to the regime of weak tunneling between the dots and the reservoirs. We propose in this paper to develop a study of the double quantum dot in the framework of the 
Keldysh non-equilibrium Green function technique (NEGF) following the same strategy as we developed\cite{Zamoum2016,Crepieux2018} before for a single quantum dot, i.e. by starting from the noninteracting case and then incorporating interactions by using 
the Keldysh NEGF technique. We present here the method and the results obtained in the case of a noninteracting double quantum dot. We give the analytical expression for the electrical current as a function of the Green functions and deduce from there the 
linear and differential conductances. In order to meet the concerns of experimentalists who have directly access to charge susceptibility via reflectometry measurements\cite{Crippa2017}, we establish the charge susceptibility of a double quantum dot related to 
the mesoscopic capacity. This study brings the foundations for further studies to come on interacting double quantum dots where the geometric configuration offers the possibility of observing Pauli spin blockade in addition to the standard Coulomb charge 
blockade.


\section{Model}

We consider two single-orbital quantum dots 1 and 2 with spin degeneracy equal to 2 ($\sigma=\pm 1$) coupled together in series through a tunnel barrier with a hopping constant $t_{\sigma}$, and connected to two metallic reservoirs $L$  and $R$ through 
spin-conserving tunnel barriers with hopping constants $t_{L\sigma}$ and $t_{R\sigma}$ respectively. In the absence of interactions, the hamiltonian writes $H = H_{DQD}+H_{leads}+H_{T}$ with
\begin{eqnarray}
H_{DQD}&=&\sum_{\sigma} \left[ \varepsilon_{1\sigma}d^{\dag}_{1\sigma}d_{1\sigma}+\varepsilon_{2\sigma}d^{\dag}_{2\sigma}d_{2\sigma} \right. \nonumber \\ &&\left. \qquad \qquad \qquad  +t_{\sigma} d^{\dag}_{2\sigma}d_{1\sigma}+t_{\sigma}^{*} d^{\dag}
_{1\sigma}d_{2\sigma} \right] \nonumber\\ 
H_{leads}&=&\sum_{k,\alpha \in (L,R),\sigma} \varepsilon_{k\alpha\sigma}c^{\dag}_{k\alpha\sigma}c_{k\alpha\sigma} \nonumber\\ 
H_{T}&=&\sum_{k\sigma} \left[ t_{L\sigma} c^{\dag}_{kL\sigma}d_{1\sigma} + t_{L\sigma}^{*} d^{\dag}_{1\sigma}c_{kL\sigma}\right. \nonumber\\ &&\left. \qquad \qquad \qquad + t_{R\sigma} c^{\dag}_{kR\sigma}d_{2\sigma} + t_{R\sigma}^{*} d^{\dag}_{2\sigma}
c_{kR\sigma} \right]
\end{eqnarray}
where $d^{\dag}_{i\sigma}$ (i=1 or 2) is the creation operator of an electron with spin $\sigma$ ($\sigma=\pm1$) in the dot i with energy $\varepsilon_{i\sigma}$; $c^{\dag}_{k\alpha\sigma}$ ($\alpha$=L or R) is the creation operator of an electron with 
momentum $k$ and spin $\sigma$ in the lead $\alpha$ with energy $\varepsilon_{k\alpha\sigma}$. The energies $\varepsilon_{i\sigma}$ in the dots are tuned by the application of a dc gate voltage $V_{Gi}$ on each dot. 
Since we are considering the noninteracting case in the absence of dot Coulomb interaction, all the results obtained in this paper are spin independent as though we were working with a spinless quantum dot system. For simplicity we will omit the
$\sigma$ subscript in the rest of the paper.

The retarded Green functions in the dots, $G^{r}_{1,1}(\omega)$, $G^{r}_{2,2}(\omega)$, $G^{r}_{1,2}(\omega)$ and $G^{r}_{2,1}(\omega)$, are solutions of the following Dyson equation written in matrix form along the $\{ 1,2 \}$ basis
\begin{eqnarray}
\bar{\bar{G}}^r(\omega) = \bar{\bar{G}}^{(0)r}(\omega)+\bar{\bar{G}}^{(0)r}(\omega)\bar{\bar{\Sigma}}^r(\omega)\bar{\bar{G}}^r(\omega)
\end{eqnarray}
where $\bar{\bar{G}}^r(\omega)$ and $\bar{\bar{G}}^{(0)r}(\omega)$ are respectively the exact and the unrenormalized Green functions in the dots, and $\bar{\bar{\Sigma}}^r(\omega)$, the  
self-energies, defined as
\begin{eqnarray}
{\bar{G}}^r(\omega)&=&
\begin{pmatrix}
G^r_{1,1}(\omega) & G^r_{1,2}(\omega) \\ G^r_{2,1}(\omega) & G^r_{2,2}(\omega)
\end{pmatrix}\nonumber\\ 
\bar{\bar{G}}^{(0)r}(\omega)&=&
\begin{pmatrix} 
(\omega - \varepsilon_1)^{-1} & 0 \\0 & (\omega - \varepsilon_2)^{-1} 
\end{pmatrix}\nonumber\\ 
\bar{\bar{\Sigma}}^r(\omega)&=&
\begin{pmatrix} 
\Sigma^r_{L}(\omega) & t^{*} \\t & \Sigma^r_{R}(\omega)
\end{pmatrix}
\label{GreenMatrix}
\end{eqnarray}
where $\Sigma^{r}_{\alpha}(\omega) = { \left| t_{\alpha} \right| }^2 \sum_k (\omega - \varepsilon_{k\alpha} + i \eta)^{-1}$ ($\eta$ being an infinitesimal positive). In the wide band limit: $\Sigma^{r}_{\alpha}(\omega)=- i \Gamma_{\alpha}(\omega)$ where $
\Gamma_{\alpha}(\omega)=\pi { \left| t_{\alpha} \right| }^2 \rho_{\alpha}^{(0)}(\omega)$ and $ \rho_{\alpha}^{(0)}(\omega)$ is the unrenormalized density of states in the reservoir $\alpha$.

By solving Eq.\eqref{GreenMatrix}, one obtains the following expressions for the Green functions in the dots 
\begin{eqnarray}
G^{r}_{1,1}(\omega)&=& \dfrac{\omega - \varepsilon_2 - \Sigma^r_{R}(\omega)}{\mathfrak{D}^r(\omega)}\nonumber\\ 
G^{r}_{1,2}(\omega)&=& \dfrac{t^{*}}{\mathfrak{D}^r(\omega)}\nonumber\\ 
G^{r}_{2,1}(\omega)&=& \dfrac{t}{\mathfrak{D}^r(\omega)}\nonumber\\ 
G^{r}_{2,2}(\omega)&=& \dfrac{\omega - \varepsilon_1 - \Sigma^r_{L}(\omega)}{\mathfrak{D}^r(\omega)}
\label{GreenFunctions}
\end{eqnarray}
with $\mathfrak{D}^r(\omega)=(\omega - \varepsilon_1 - \Sigma^r_{L}(\omega))(\omega - \varepsilon_2 - \Sigma^r_{R}(\omega)) -  { \left| t \right| }^2$.



\section{General expression for the current}

We derive the expression for the current through the double quantum dot by using the Keldysh nonequilibrium Green function technique.

The current $I_{L}$ from the $L$ reservoir to the central region can be calculated from the time evolution of the occupation number operator in the $L$ reservoir
\begin{eqnarray}
I_{L} = -e \left\langle \dfrac{d\hat{n}_{L}(t)}{dt} \right\rangle= -ie \left\langle \left[\hat{H},\hat{n}_{L}\right] \right\rangle
\end{eqnarray}
where $\hat{n}_{L}(t)=\exp(i\hat{H}t) \hat{n}_{L} \exp(-i\hat{H}t)$ is the number of electrons in the L reservoir in the Heisenberg representation (with $\hat{n}_{L}=\sum_{k,\alpha \in (L)} c^{\dag}_{k\alpha}c_{k
\alpha}$). The current $I_{R}$ from the $R$ reservoir to the central region can be defined in an analogous way.

Defining the lesser Green functions mixing the electrons in the dot and in the reservoir according to $G^{<}_{k\alpha, i}(t,t')=i \left\langle d^{\dag}_{i}(t') c_{k\alpha}(t)  \right\rangle$ and $G^{<}_{i,k\alpha}(t,t')=i \left\langle 
c^{\dag}_{k\alpha}(t') d_{i}(t)  \right\rangle$, the currents write
\begin{eqnarray}
&&I_{L} = e \sum_{k} \left[ t_{L} G^{<}_{1,kL} (t,t) - t^{*}_{L} G^{<}_{kL,1} (t,t) \right]\label{ILsigEXACT}
\end{eqnarray}
and a similar expression for $I_{R}$. The lesser Green functions $G^{<}_{k\alpha, i}(t,t')$ and $G^{<}_{i,k\alpha}(t,t')$ are then evaluated by applying the analytic continuation rules provided by the Langreth theorem \cite{Haug2008} to the Dyson equations for the 
Green functions. It results in
\begin{eqnarray}
&&I_{L} = i \dfrac{e}{\pi}  \int_{-\infty}^{\infty} d\omega \Gamma_{L}(\omega) \nonumber \\ &&\times \left[ G^{<}_{1,1} (\omega) + n_F^L(\omega) \left(G^{r}_{1,1} (\omega)  - G^{a}_{1,1} (\omega) \right)  \right] 
\end{eqnarray}
where $n_F^\alpha(\omega)=\frac{1}{e^{\beta(\omega-\mu_{\alpha})}+1}$ is the Fermi-Dirac distribution function in the reservoir $\alpha$ with chemical potential $\mu_{\alpha}$.

When the system is in the steady state, one gets: 
\begin{eqnarray}
I_{L} = \dfrac{2e}{\pi}  \int_{-\infty}^{\infty} d\omega && \Gamma_{L}(\omega) G^r_{1,2}(\omega)\Gamma_{R}(\omega)G^a_{2,1}(\omega)\nonumber\\ &&\times \left[n_F^L(\omega)-n_F^R(\omega)\right]
\label{ILsigSANSINTERACTION}
\end{eqnarray}
and $I_{L} = - I_{R}$, where $G^{a}_{i,j}(\omega)$ are the four advanced Green functions in the dots.

\section{Charge susceptibility}

In order to calculate the charge susceptibility of the system, one needs to connect each quantum dot $i$ through a capacitance $C_{ac}^{i}$ to an ac voltage $V_{ac}(t)$ (see Ref. \onlinecite{Cottet2011}), bringing the additional following term to the hamiltonian 
$\hat{H}$: 
$H_{ac}(t) =\sum_{i=1,2} e\alpha_{i}\hat{n}_{di}V_{ac}(t)$ where $\hat{n}_{i}=d_{i}^{\dag}d_{i}$, the number of electrons in the dot i, and $\alpha_{i}$ measures the charge on 
the capacitance $C_{ac}^{i}$. 

The total charge $\hat{Q}_{ac}$ on the capacitances is given by
\begin{eqnarray}
\hat{Q}_{ac} =\sum_{i=1,2} \left[ -e\alpha_{i}\hat{n}_{di}eV_{ac}(t) \right] +(C_{1}^{(0)}+C_{2}^{(0)}) V_{ac} 
\label{Q_ac}
\end{eqnarray}
where $C_{1}^{(0)}$ and $C_{2}^{(0)}$ are the capacitances of the two quantum dots when they are isolated (i.e. for $t=t_{L}=t_{R}=0$).

From Eq.\eqref{Q_ac} and by using the linear response theory, one can obtain the charge susceptibility $\chi(t-t')$ 
\begin{eqnarray}
\chi_{\sigma}(t-t')=-i \theta(t'-t) \sum_{ij} \alpha_{i}\alpha_{j} \left\langle \left[ \hat{n}_{i\sigma}(t'),\hat{n}_{j\sigma}(t) \right] \right\rangle
\end{eqnarray}

By taking its Fourier transform, one gets the dynamical charge susceptibility $\chi(\omega)$ and in particular the static charge susceptibility in the $\omega=0$ limit. The static charge susceptibility can simply be derived from
\begin{eqnarray}
\chi(\omega=0)=\sum_{i,j}\alpha_i \alpha_j \dfrac{\partial \langle \hat{n}_{i} \rangle_0}{\partial \varepsilon_{j}}
\label{static}
\end{eqnarray}
where  $\langle \hat{n}_{i} \rangle_0$ is the expectation value of the occupancy in the dot $i$ at $V_{ac}(t)=0$, which can be calculated from the lesser Green functions by using: $\langle \hat{n}_{i} \rangle=-\dfrac{i}{2\pi}\int d\omega G^{<}_{i,i}
(\omega)$. In the case when both $\Gamma_{\alpha}(\omega)$ is independent on $\omega$, it is straightforward to calculate $\langle \hat{n}_{i} \rangle_0$ and then take its derivative with respects to $\varepsilon_{j}$ 
which allows to find the charge susceptibility 
$\chi(\omega=0)$.

\section{Results}

\begin{figure}[ht]
\centering
\includegraphics[width=8.5cm]{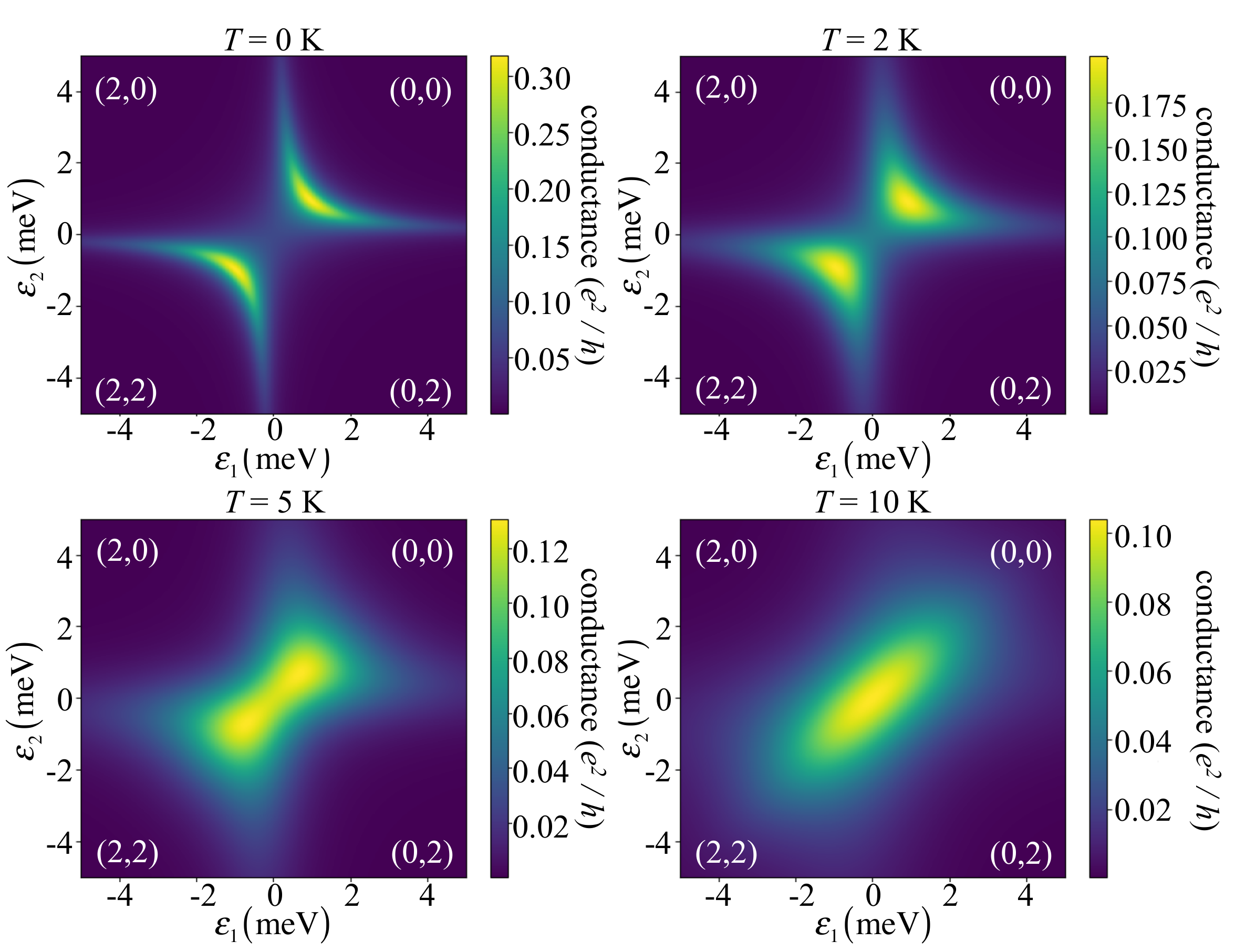}
\caption{Color-scale plots of the linear conductance $G$ of the noninteracting double quantum dot as a function of the energy levels $\varepsilon_{1}$ and $\varepsilon_{2}$ in the dots for $\Gamma_{L}~=~\Gamma_{R}~=~0.25~
\mathrm{meV}$ (symmetric couplings), $t=1~\mathrm{meV}$ and $\mu_L=\mu_R=0$ at four different temperatures $T = 0, 2, 5, 10$~K. $(n_1, n_2)$ denotes the state of the system with occupation numbers $n_1$ and $n_2$ in each dot.}
\label{fig2}
\end{figure}

\begin{figure}[ht]
\centering
\includegraphics[width=7.5cm]{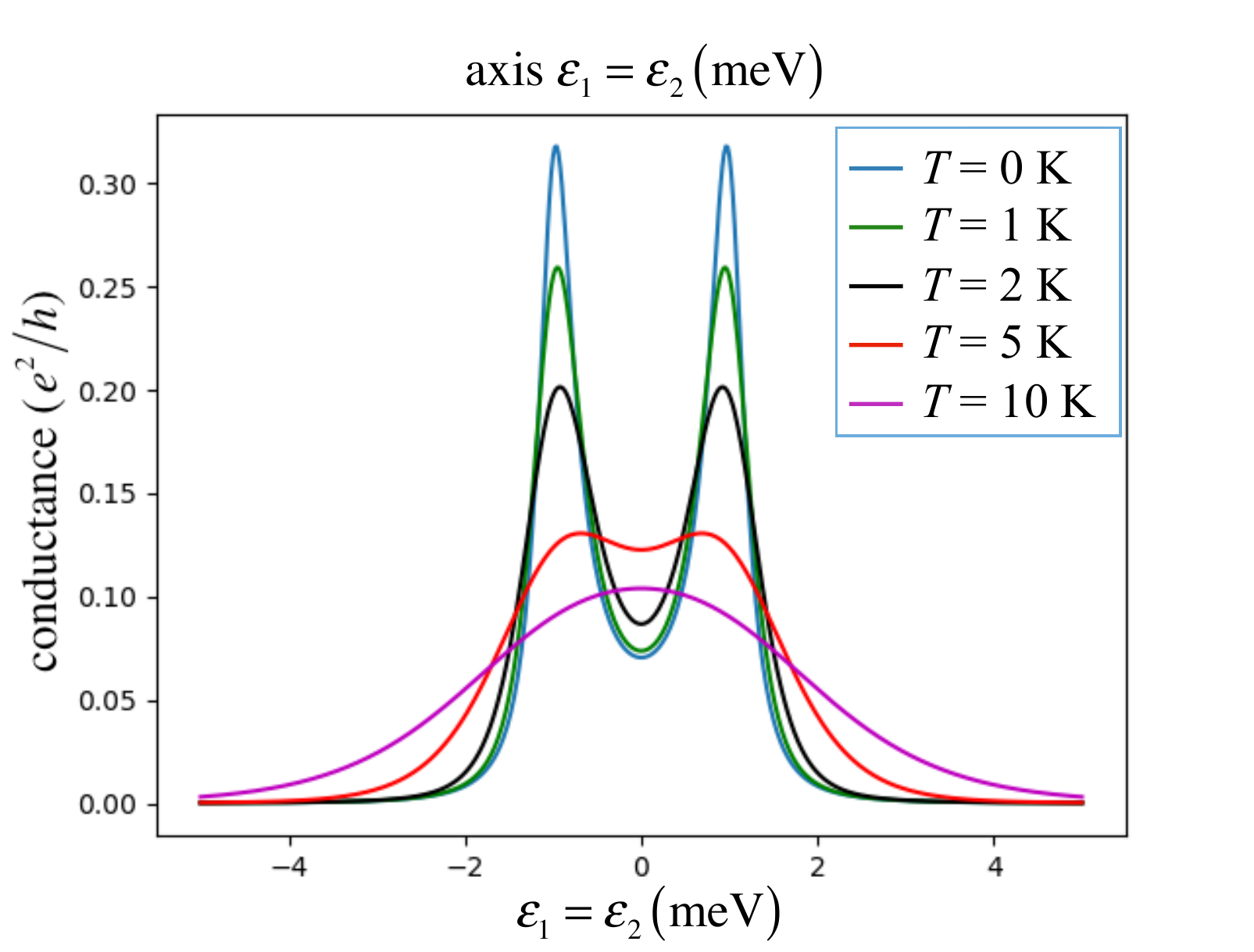}
\caption{Linear conductance $G$ as a function of the energy $\varepsilon_{1}$ along the first diagonal $\varepsilon_{1}=\varepsilon_{2}$ of the plots in \figurename{\ref{fig2}} at five different temperatures $T = 0, 1, 2, 5, 10$~K.}
\label{fig3}
\end{figure}

The color-scale plots of the linear conductance are shown in \figurename{\ref{fig2}} as a function of the energy levels $\varepsilon_{1}$ and $\varepsilon_{2}$ in the dots for $\mu_L=\mu_R=0$ at four different temperatures. \figurename{\ref{fig3}} reports the 
dependence of $G$ with the energy $\varepsilon_{1}$ along the first diagonal $\varepsilon_{1}=\varepsilon_{2}$ of the previous figure. The state of the system with occupation numbers $n_1$ and $n_2$ in each dot is denoted as $(n_1, n_2)$. At low 
temperature, the states $(0,0)$ and $(2,2)$ are clearly separated from the $(0,2)$ and $(2,0)$ states by two conductance peaks thanks to the effect of the finite interdot hopping term $t$. With increasing temperatures, this frontier is getting blurrier and the 
conductance is higher along the $(0,2)$-$(2,0)$ frontier. 

\begin{figure}[h]
\centering
\includegraphics[width=8.5cm]{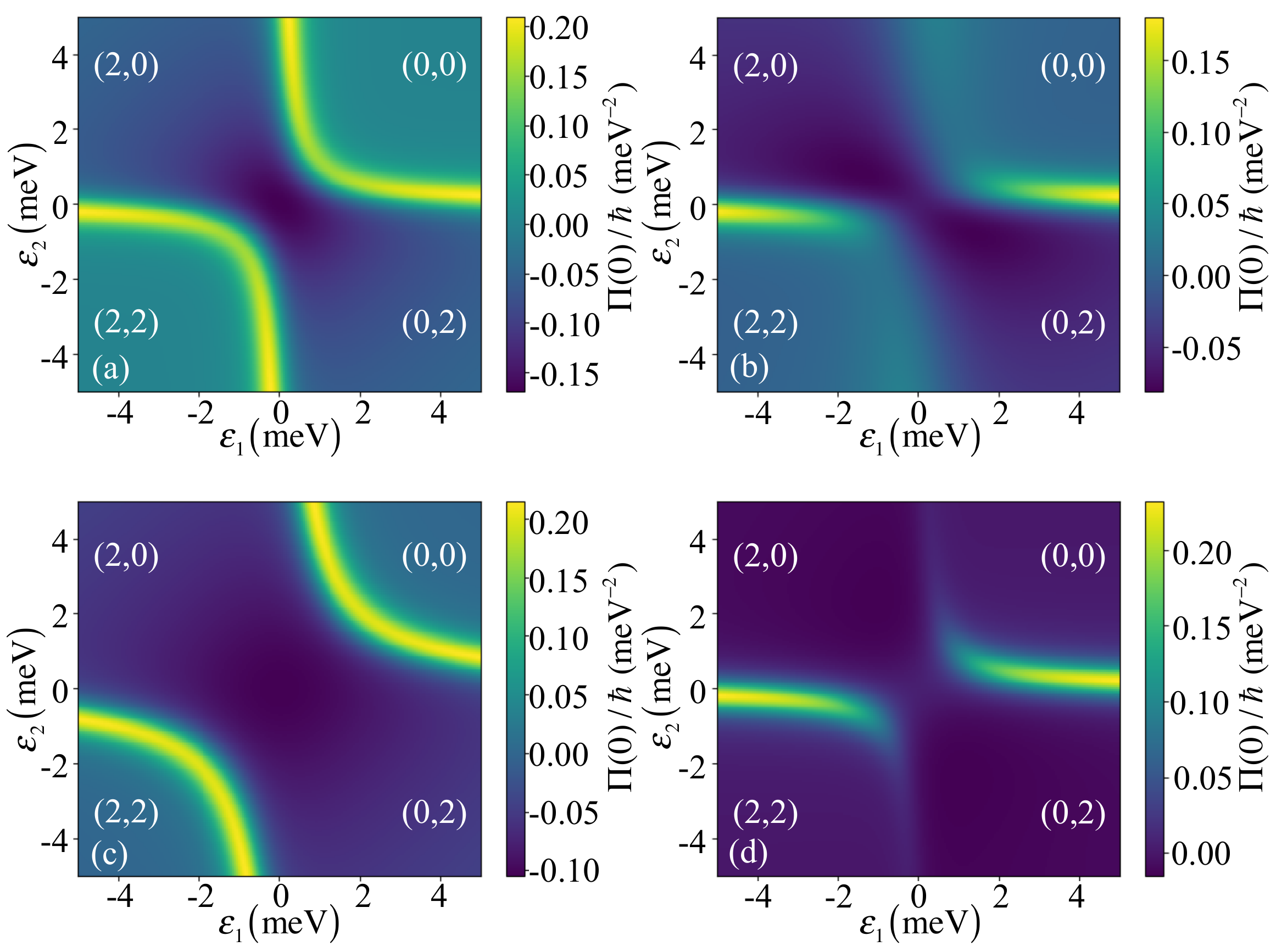}
\caption{Color-scale plots of the static charge susceptibility $\chi(0)$ of the noninteracting double quantum dot as a function of the energy levels $\varepsilon_{1\sigma}$ and $\varepsilon_{2\sigma}$ in the dots at $T=1$~K and $
\mu_L=\mu_R=0$ for four sets of parameters (a) $\Gamma_{L\sigma}=\Gamma_{R\sigma}= 0.25$~meV (symmetric couplings), $\alpha_L = \alpha_R = -0.5$ (symmetric geometry), $t_{\sigma}= 1$~meV ; (b) $\Gamma_{L\sigma}= 5\Gamma_{R\sigma} 
1.25$~meV (asymmetric couplings), $\alpha_L = \alpha_R= -0.5$ (symmetric geometry), $t_{\sigma} = 1$~meV ; (c) $\Gamma_{L\sigma}= \Gamma_{R\sigma} = 0.25$~meV (symmetric couplings), $\alpha_L = \alpha_R = -0.5$ (symmetric geometry), 
$t_{\sigma} = 2$~meV~; and (d) $\Gamma_{L\sigma} = \Gamma_{R\sigma} = 0.25$~meV (symmetric couplings), $\alpha_R = 5\alpha_L = -0.5$ (asymmetric geometry); $t_{\sigma}= 1$~meV.}
\label{fig4}
\end{figure}

\figurename{\ref{fig4}} shows the static charge susceptibility $\chi(\omega)$ at $T=1$~K and $\mu_L=\mu_R=0$ for four different configurations of couplings to the reservoirs, capacitances $\alpha_L, \alpha_R$ (related to the geometry 
of the device) and interdot hopping $t$. 
It shows the existence of peaks for the static charge susceptibility in the $(\varepsilon_1,\varepsilon_2)$ plane along two arcs located in the 1st and 3rd quadrants ($\varepsilon_1 >0$, $\varepsilon_2 >0$, and $\varepsilon_1 <0$, 
$\varepsilon_2 <0$ respectively). The corner spaces encircled by these two arcs correspond to the regimes (0,0) and (2,2) respectively. The central region between the two arcs corresponds to the other two regimes (0,2) and (2,0). As can be seen, the charge susceptibilities are equal in both (0,0) and (2,2) regimes, but differ from the one observed in the (0,2) and (2,0) regimes. This can be easily understood on the basis of the following physical 
argument. Let us first point out that in the limit $t\ll (\Gamma_L, \Gamma_R$), the peaks in $\chi(\omega)$ occur near the two horizontal and vertical axes delimiting the four (0,0), (2,0), (2,2), (0,2) regimes, with an equal $\chi(\omega)$ in each quadrant brought 
by the intradot transition contributions only (i.e. by the $\dfrac{\partial \langle \hat{n}_{i} \rangle_0}{\partial \varepsilon_{i}}$ terms). In the presence of a finite $t$, the latter pattern transforms into two arcs located in the 1st and 3rd quadrants 
respectively as mentioned above. The larger $t$ is, the larger the distance between the two arcs is, as can be seen by comparing \figurename{\ref{fig4}} a and c. Gradually as the two arcs are formed from the initial pattern, the contributions to the charge 
susceptibility brought by the interdot transitions (i.e.$\dfrac{\partial \langle \hat{n}_{i} \rangle_0}{\partial \varepsilon_{j}}$ with $i \neq j$) become more and more important, showing a strong dependence inside the ($\varepsilon_1,\varepsilon_2$) plane. 
Consequently, $\chi(\omega)$ in the (0,0) and (2,2) regimes belonging to the two quadrants inside which the arcs are formed, differ from $\chi(\omega)$ in the other (0,2) and (2,0) regimes, which explains the difference observed in \figurename{\ref{fig4}}. The 
last comment concerns the role of an asymmetry in either dot-lead couplings or geometry of the device. As can be seen, the effect of an asymmetry is to reduces the intensity of $\chi(\omega)$ along one of the arms of the arcs.

\section{Conclusion}

We have studied the linear and differential conductances as well as the charge susceptibility of a noninteracting quantum dot by using the Keldysh nonequilibrium Green function technique. The obtained expressions are exact and allows one to study the 
variation of the conductances and charge susceptibility with temperature and any parameters of the double quantum dot model, energy levels $\varepsilon_{1}, \varepsilon_{2}$ of the dots, $\Gamma_{L}$, $\Gamma_{R}$ and 
interdot hopping $t$. We have then discussed the evolution of the stability diagram of the system with the different parameters. This work opens the way for extension to the case of a double quantum dot in the presence of Coulomb 
interactions as is relevant for spin-qubit silicon-based devices.

{\it Acknowledgments} -- For financial support, the authors acknowledge the Programme Transversal Nanosciences of the CEA and the CEA Eurotalents Program. 

\section*{Bibliography}
\bibliography{DQD}

\end{document}